\documentclass[aps,prb,twocolumn,a4paper,superscriptaddress,showpacs]{revtex4-1}

	\usepackage{graphicx}
	\usepackage{dcolumn}
	\usepackage{bm}
	\usepackage{color}

	\bibpunct{[}{]}{,}{n}{}{}	
	\newcommand{\sn}{Co$_{1.21}$V$_{1.79}$O$_{4}$} 
	\definecolor{dgreen}{rgb}{0,0.7,0}


\begin{document}
\title{Electronic States and Possible Origin of the Orbital-Glass State in a Nearly Metallic Spinel Cobalt Vanadate: An X-ray Magnetic Circular Dichroism Study}

	\author{Yosuke Nonaka}
	\affiliation{Department of Physics, the University of Tokyo, Bunkyo-ku, Tokyo 113-0033, Japan}
	\email{nonaka@wyvern.phys.s.u-tokyo.ac.jp}
	\author{Goro Shibata}
	\affiliation{Department of Physics, the University of Tokyo, Bunkyo-ku, Tokyo 113-0033, Japan}
	\author{Rui Koborinai}
	\affiliation{Department of Physics, Waseda University, Tokyo 169-8555, Japan}
	\author{Keisuke Ishigami}
	\affiliation{Department of Physics, the University of Tokyo, Bunkyo-ku, Tokyo 113-0033, Japan}
	\author{Shoya Sakamoto}
	\affiliation{Department of Physics, the University of Tokyo, Bunkyo-ku, Tokyo 113-0033, Japan}
	\author{Keisuke Ikeda}
	\affiliation{Department of Physics, the University of Tokyo, Bunkyo-ku, Tokyo 113-0033, Japan}
	\author{Zhendong Chi}
	\affiliation{Department of Physics, the University of Tokyo, Bunkyo-ku, Tokyo 113-0033, Japan}
	\author{Tsuneharu Koide}
	\affiliation{Photon Factory, Institute of Materials Structure Science, High Energy Accelerator Research Organization (KEK), Tsukuba, Ibaraki 305-0801, Japan}
	\author{Arata Tanaka}
	\affiliation{Department of Quantum Matter, ADSM, Hiroshima University, Higashi-Hiroshima 739-8530, Japan}
	\author{Takuro Katsufuji}
	\affiliation{Department of Physics, Waseda University, Tokyo 169-8555, Japan}
	\affiliation{Kagami Memorial Research Institute for Materials Science and Technology, Waseda University, Tokyo 169-0051, Japan}
	\author{Atsushi Fujimori}
	\affiliation{Department of Physics, the University of Tokyo, Bunkyo-ku, Tokyo 113-0033, Japan}

\date{\today}

\begin{abstract}
	We have investigated the orbital states of the orbital-glassy (short-range orbital ordered) spinel vanadate \sn\ using x-ray absorption spectroscopy (XAS), x-ray magnetic circular dichroism (XMCD), and subsequent configuration-interaction cluster-model calculation. 
	From the sign of the XMCD spectra, it was found that the spin magnetic moment of the Co ion is aligned parallel to the applied magnetic field and that of the V ion anti-parallel to it, consistent with neutron scattering studies. 
	It was revealed that the excess Co ions at the octahedral site take the trivalent low-spin state, and induce a random potential to the V sublattice. 
	The orbital magnetic moment of the V ion is small, suggesting that the ordered orbitals mainly consists of real-number orbitals.
\end{abstract}

\maketitle

\section{Introduction}
\begin{figure}
		\includegraphics[width=8.6cm]{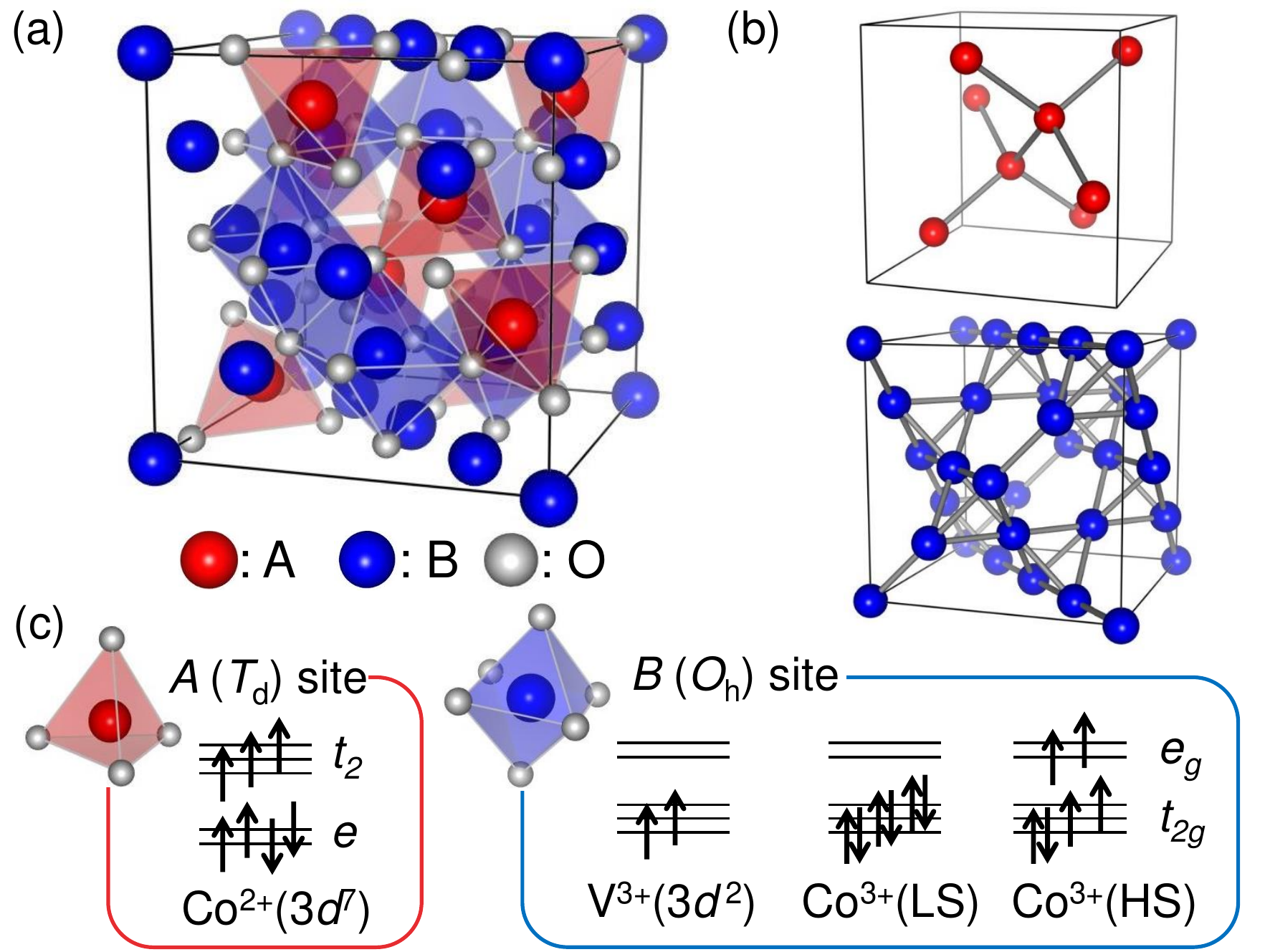}
		\caption{
			Illustrations of the crystal structure and the electronic configurations of the $3d$ orbitals of spinel-type oxides. 
			(a) Cubic unit cell of \textit{AB}$_2$O$_4$. The red (blue) polyhedra denote the \textit{A}O$_4$ tetrahedra (\textit{B}O$_6$ octahedra). 
			(b) Pyrochlore and diamond structures consisting of the \textit{B} and \textit{A} site network, respectively. 
			(c) Electronic configuration of the Co$^{2+}$ ion at the tetrahedral (A) site and those of the V$^{3+}$ and Co$^{3+}$ ions at the octahedral (B) site. 
		}\label{fig:structure}
	\end{figure}

	Spinel-type transition-metal oxides \textit{AB}$_2$O$_4$ [Fig. \ref{fig:structure} (a)] are one of the most attractive playgrounds for the studies of emergent physical properties arising from multiple degrees of freedom in strongly correlated electron systems. 
	This structure consists of the diamond sublattice of the \textit{A} site and the pyrochlore sublattice of the \textit{B} site [Fig. \ref{fig:structure} (b)]. 
	Particularly in the pyrochlore sublattice, there is strong geometrical spin frustration. 
	In the spinel-type vanadates \textit{A}V$_2$O$_4$, in addition to the spin frustration, the orbital degree of freedom also exists in the triply degenerate V $t_{2g}$ orbitals, as shown in Fig. \ref{fig:structure} (c).
	Therefore, a number of experimental and theoretical studies have been performed on the spinel-type vanadates, particularly on the orbital ordering in $A$V$_2$O$_4$, where $A$ = Mg \cite{Wheeler2010}, Mn \cite{Suzuki2007, Garlea2008, Sarkar2009a, Nii2012, Nii2013, Matsuura2015, Okabayashi2015}, Fe \cite{Kang2012, Nii2012,Nii2013,Kawaguchi2016a}, Zn \cite{Maitra2007,Lee2004}, and Cd \cite{Nishiguchi2002,Onoda2003}. 
	All of them exhibit structural phase transitions from the cubic to a low-symmetry phase with decreasing temperature, and the orbital ordering is usually associated with the structural phase transition. 

	To understand the orbital ordering in these systems, theoretical models based on the pyrochlore network of V ions have been studied \cite{Tsunetsugu2003,Motome2004,Khomskii2005,Tchernyshyov2004}. 
	Through a strong coupling approach using the Kugel-Khomskii-type Hamiltonian \cite{Kugel1982}, Tsunetsugu and Motome proposed an antiferro-orbital ordering model for ZnV$_2$O$_4$ \cite{Tsunetsugu2003,Motome2004}, which consists of the alternating $d_{zx}$ and $d_{yz}$ orbitals. 
	Khomskii and Mizokawa proposed an orbital-Peierls model \cite{Khomskii2005}, which is induced by the dimerization of orbitals in 1D chains consisting of the $\sigma$-bonds of the $d_{zx}$ and $d_{yz}$ orbitals. 
	In contrast to these real orbital models, Tchernyshyov proposed a ferro-orbital model, where the $d_{xy}$ and one of the complex $d_{zx} \pm id_{yz}$ orbitals are occupied by two electrons \cite{Tchernyshyov2004}. 
	This complex-number orbital have a large orbital magnetic moment ($\pm 1 \mu_B$ at most). 
	In addition to the above orbital ordering mechanisms, a spontaneous lattice-distortion mechanism without orbital ordering was also proposed by Pardo \textit{et al.,} \cite{Pardo2008} for nearly metallic spinel-type vanadate, where partially delocalized V 3$d$ electrons drive a homopolar bond formation and V-V dimerization in a system where the $U/W$ (the ratio of the on-site Coulomb energy to the bandwidth) is not sufficiently larger than one.

	Among the spinel-type vanadates, CoV$_2$O$_4$ has attracted particular attention because of the close proximity to the Mott-type metal-insulator-transition from the insulator side due to the short V-V bond length \cite{Kismarahardja2011,Kiswandhi2011}. 
	This compound shows ferrimagnetism below $T_{\rm N}$ = 160 K, but a structural phase transition nor a signature of orbital ordering had not been seen down to 10 K by x-ray diffraction \cite{Kismarahardja2011}. 
	The absence of orbital ordering is consistent with the itinerant character of V $3d$ electrons in CoV$_2$O$_4$. 
	The absence of orbital ordering was also proposed by extrapolating results of neutron scattering study on Mn$_{1-y}$Co$_y$V$_2$O$_4$ ($y \leq 0.8$) single crystals to $y = 1.0$ \cite{Ma2015}. 
	However, recently a very small structural phase transition ($\Delta a/a \sim 10^{-4}$) at 90 K was reported by a neutron scattering study of stoichiometric CoV$_2$O$_4$ powders \cite{Reig-i-Plessis2016}. 
	A small distortion was also reported for single crystals of Co$_{1+x}$V$_{2-x}$O$_4$ ($0.16 \leq x \leq 0.3$) by strain gauge measurements, and an orbital-glass (short-range orbital ordering) state was proposed from its gradual lattice distortion and the slow dynamics of the dielectric constant \cite{Koborinai2016}. 
	Very recently, two structural phase transitions, one at 95 K [cubic to tetragonal ($I4_1/amd$)] and the other at 59 K [tetragonal ($I4_1/amd$) to tetragonal ($I4_1/a$)], were reported by neutron powder diffraction and single-crystal synchrotron-radiation x-ray diffraction measurements on stoichiometric CoV$_2$O$_4$ \cite{Ishibashi2017}.
	It was also proposed that a long-range antiferro-orbital ordering exists below the 59-K structural phase transition, where an anomaly in the specific heat was reported \cite{Huang2012,Shimono2016}.
	Since the anomaly persists up to $x \sim 0.15$ \cite{Shimono2016}, the long-range orbital-ordering should also persist up to $x \sim 0.15$.
	On the other hand, because of the close proximity of CoV$_2$O$_4$ to the Mott-type metal-insulator transition, the homopolar bond formation mechanism \cite{Pardo2008} might be the origin of the lattice distortion. 

	In the present study, in order to obtain more direct information about the V $3d$ electrons, we have measured $L$-edge x-ray absorption spectra (XAS) and x-ray magnetic circular dichroism (XMCD) of CoV$_2$O$_4$.
	Because $L$-edge XAS is a well established probe of the $3d$ electronic structure, and $L$-edge XMCD gives us the quantitative information about the orbital magnetic moments which is the key information about the orbital states.
	It should be noted that high-quality single crystals of CoV$_2$O$_4$ can be grown only for compositions with excess Co.
	Therefore, the composition of the crystals studied in this work was \sn\ \cite{Koborinai2016}.
	The excess Co ions enter the octahedral site and significantly affect the physical properties of Co$_{1+x}$V$_{2-x}$O$_4$. 
	Upon increasing excess Co, the activation energy for the transport becomes higher \cite{Rogers1963,Rogers1966}, $T_{\rm N}$ becomes higher, and the onset temperature of the orbital-glass state becomes lower \cite{Koborinai2016}. 
	Therefore, it is necessary to elucidate the electronic state of excess Co ions, and $L$-edge XAS and XMCD are ideal tools for that purpose, too. 
	Using the configuration-interaction (CI) cluster-model analysis of the XAS and XMCD data, one can discuss the electronic states of the Co ions at the tetrahedral site and the excess Co ions at the octahedral site separately.

\section{Methods of Experiment and Calculation}
	\begin{figure}
		\includegraphics[width=8.6cm]{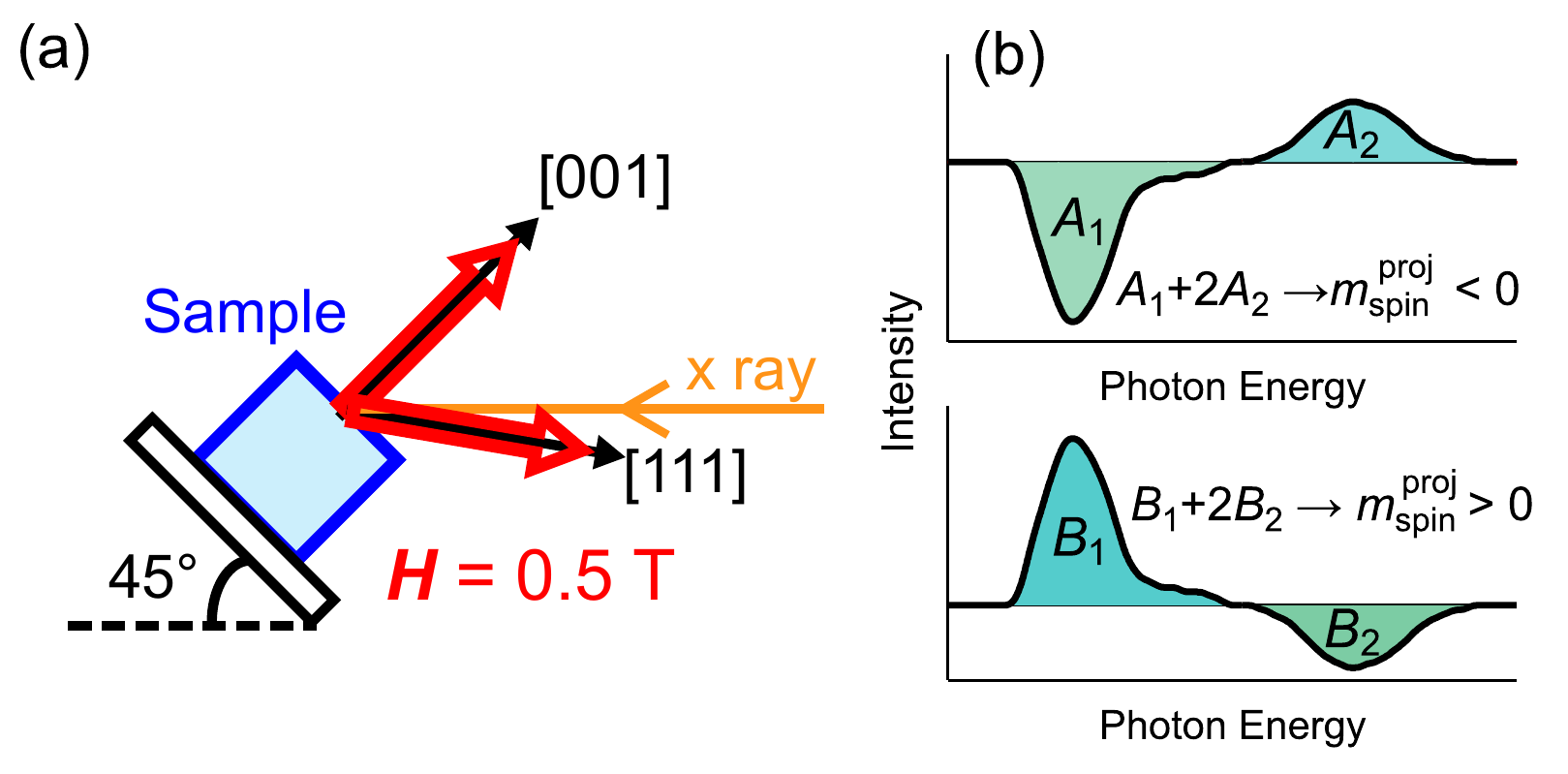}%
		\caption{
			Schematic pictures of the experimental geometry (a) and the sign of XMCD signals (b). 
			The upper and lower panels of (b) show typical XMCD spectra where the projection of the spin magnetic moment onto the X-ray incident vector $m_{\rm spin}^{\rm proj}$ is negative and positive, respectively.
		}\label{fig:geo}
	\end{figure}
	Bulk single crystals of \sn\ were grown by the floating zone method. 
	In order to avoid the formation of the V$_2$O$_3$ impurity phase, the compositions of the feed rod was not stoichiometric ($\text{Co}:\text{V} = 1:2$) but containing 50\% excess Co ($\text{Co}:\text{V} = 1.5:2$) relative to the stoichiometric one. 
	The $\text{Co}:\text{V}$ ratio of the synthesized single crystals was determined to be $1.21:1.79$ by induction-coupled plasma (ICP) analysis \cite{Koborinai2016}.

	Magnetic field angle-dependent XAS and XMCD were measured at the undulator beamline BL-16A of Photon Factory. 
	The spectra were taken in the total electron-yield (TEY) mode at 30, 70, and 110 K. 
	The sample angle was fixed so that the incident x-rays form an angle of \(45^\circ\) from the [001] direction and within the [001]--[111] plane in the cubic notation. (Hereafter, we use the cubic notation.) 
	A magnetic field of 0.5 T was applied along the [001] and [111] directions using a `vector magnet' XMCD apparatus \cite{Furuse2013}. 
	A schematic picture of the experimental geometry is shown in Fig. \ref{fig:geo} (a).
	In order to obtain clean surfaces, the samples were cleaved in vacuum (base pressure $< 8 \times 10^{-10}$ Torr). 
	The XMCD intensities have been corrected for the degree of circular polarization of BL-16A.

	\begin{figure*}
		\includegraphics[width=15cm]{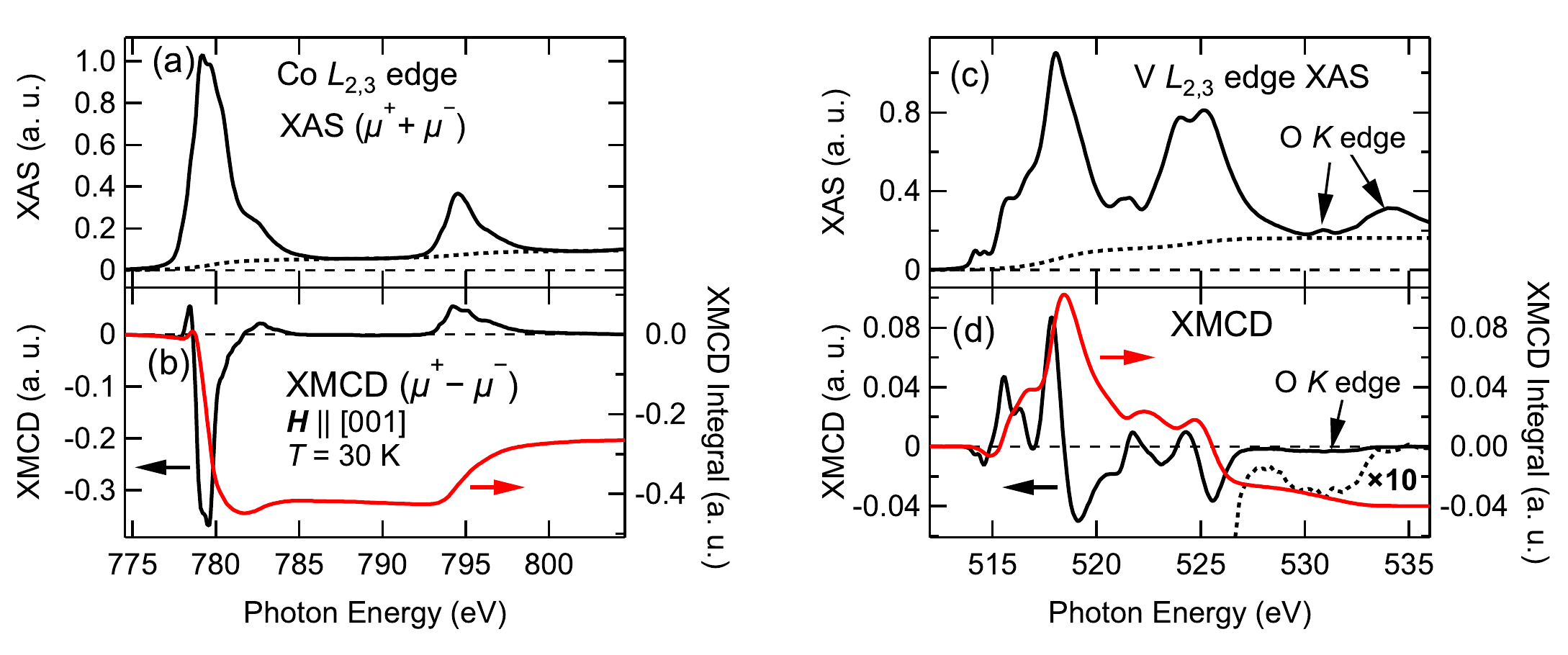}%
		\caption{
			XAS and XMCD spectra of \sn. 
			(a), (b) XAS and XMCD at the Co $L_{2,3}$ edge. 
			(c), (d) XAS and XMCD at the V $L_{2,3}$ edge. 
			The dotted curves in (a) and (c) are two-step background functions representing the edge jump. 
			The dotted curves in (d) is a magnified XMCD spectrum around 531 eV.
		}\label{fig:exp}
	\end{figure*}
	
	To examine the detailed electronic structure, we have fitted the CI cluster-model calculation to the experimental XAS and XMCD spectra \cite{Tanaka1994}.
	In the calculation, we adopted the empirical ratio between the on-site Coulomb energy $U_{dd}$ and the $3d$--$2p$ hole Coulomb energy $U_{dc}$, $U_{dc}/U_{dd} \sim 1.3$ \cite{Bocquet1992,Hufner2003}, and that between Slater-Koster parameters ($pd\sigma$) and ($pd\pi$), ($pd\sigma$)/($pd\pi$)$ \sim -2.17$ \cite{Harrison1980}. 
	Hybridization strength between the O $2p$ orbitals $T_{pp}$ was fixed at 0.7 eV (for the $O_h$ site) and 0 eV (for the $T_d$ site) \cite{Tanaka1994,Chen2004}. 
	The ionic Hartree-Fock (HF) values were used for the spin-orbit interaction constants of Co 3$d$, Co 2$p$, and V 2$p$ states, and	80\% of the ionic HF values were used for the Slater integrals \cite{Tanaka1994}. 
	Thus, the crystal-field splitting 10$Dq$, $U_{dd}$, the charge-transfer energy $\it{\Delta}$, and ($pd\sigma$) were treated as fitting parameters. 
	For each absorption edge, further assumptions (as described) were adopted to reduce the number of adjustable parameters.

\section{Results and Discussion}
	Figure \ref{fig:exp} shows the XAS ($\mu^++\mu^-$) and XMCD ($\mu^+-\mu^-$) spectra of the \sn\ single crystal at the V and Co $L_{2, 3}$ edges under the magnetic field of 0.5 T and at the temperature of 30 K.
	Here, $\mu^+$ ($\mu^-$) denote the absorption coefficients for the photon helicities parallel (antiparallel) to the majority spin direction.
	The energy-integrated XMCD spectra are also shown in Figs. \ref{fig:exp} (b) (d) by red solid curves. 
	The sign of XMCD provides us with the information about the spin magnetic moment. 
	As shown in Fig. \ref{fig:geo} (b), the sign of XMCD at the $L_2$ and $L_3$ edges indicates whether the projection of the spin magnetic moment onto the X-ray incident vector $m_{\rm spin}^{\rm proj}$  is positive or negative. 
	According to the ionic picture shown in Fig. \ref{fig:structure} (c), since the total spin magnetic moment of the two V$^{3+}$ ($2\mu_B \times 2$ ions $= 4\mu_B$) ions in the unit formula is larger than that of the Co$^{2+}$ ($3\mu_B$) ion, the spin magnetic moment of the V$^{3+}$ ion is expected to be parallel to the applied magnetic field and that of the Co$^{3+}$ ion antiparallel to it. 
	However, the sign of the XMCD signals in the Fig. \ref{fig:exp} (b) and (d) shows that the magnetic moment of the Co ion and of the V ion are, respectively, parallel and anti-parallel to the magnetic field. 
	This spin orientation and the reduction of the spin magnetic moment of V from the ionic value is consistent with previous neutron studies on Mn$_{1-y}$Co$_y$V$_2$O$_4$ and Co$_{1+x}$V$_{2-x}$O$_4$ \cite{Ma2015,Koborinai2016,Reig-i-Plessis2016}, reflecting the strong geometrical frustration of the pyrochlore-type V sublattice. 
	The XAS and XMCD spectra at both edges show clear multiplet structures, reflecting that the V and Co electrons of this system are still localized even though the itineracy is indeed increased due to the short V--V bond length \cite{Kismarahardja2011,Kiswandhi2011}.
	The spectral line shapes at the V edge is very similar to those of other ferrimagnetic insulating $A$V$_2$O$_4$ ($A$=Fe, Mn) \cite{Kang2012,Matsuura2015,Okabayashi2015}. 
	This similarity also indicates that the V $3d$ electrons of \sn\ are almost localized similarly to the insulating $A$V$_2$O$_4$ ($A$=Fe, Mn). 
	Therefore the homopolar bond formation mechanism \cite{Pardo2008} which cannot take place for well localized V 3$d$ electrons, where the $U/W$ ratio should be large, may be ruled out as the origin of the lattice distortion in this material.
	It should be noted that the other ferrimagnetic insulating $A$V$_2$O$_4$ ($A$=Fe, Mn) show clearer structural phase transitions and are more distorted than Co$_{1+x}$V$_{2-x}$O$_4$. In particular, for MnV$_2$O$_4$, $(c/a)-1$ in the low-temperature tetragonal phase is 50 times larger than those of Co$_{1+x}$V$_{2-x}$O$_4$ \cite{Koborinai2016}, indicating that the tetragonal distortions (elongation or contraction along the $\langle001\rangle$ directions), which accompany the orbital ordering, do not have a significant effect on the spectral line shapes of XAS and XMCD. 
	A weak but clear XMCD signal was observed around the small O \textit{K}-edge XAS pre-peak around 531 eV (Fig. \ref{fig:exp} (d)) similarly to the reports on manganese oxides \cite{Koide2001, Koide2014}. 
	This XAS pre-peak is considered to arise from the 3\textit{d}--2\textit{p} hybridization.
	The observed XMCD signal should originate from the O 2\textit{p}-hole orbital polarization induced by the overlap of the 3\textit{d} and 2\textit{p} orbital wave functions.
	According to the XMCD orbital sum rule \cite{Thole1992}, the integral of the XMCD spectrum is proportional to the orbital magnetic moment. 
	Therefore, the XMCD data show that the orbital magnetic moment of O is parallel to the spin and orbital magnetic moments of Co and parallel to the orbital magnetic moment of V.

	\begin{figure}
		\includegraphics[clip, width=8.6cm]{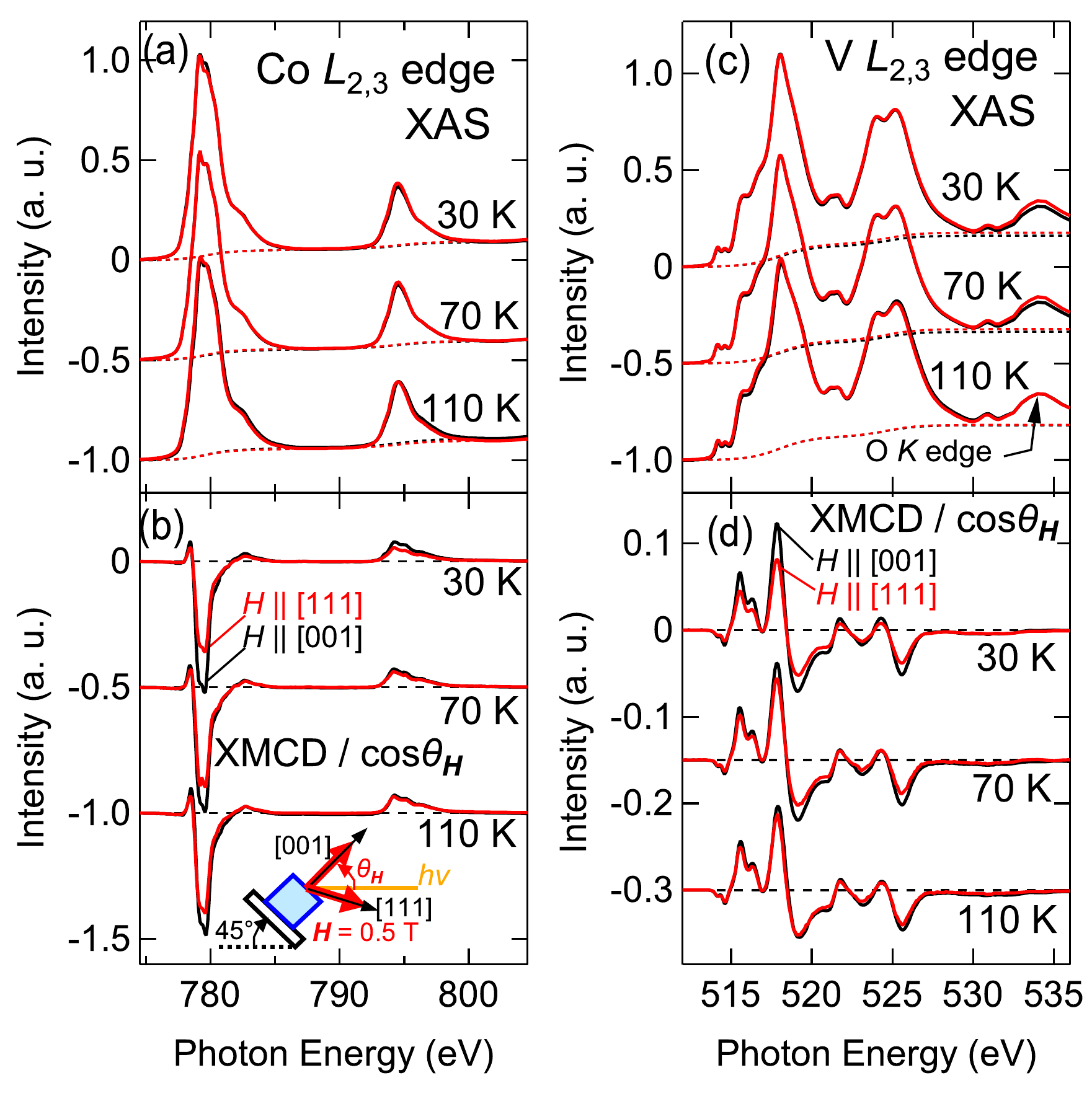}
		\caption{
			Experimental spectra of \sn\ under the magnetic field of 0.5 T along the [001] and [111] directions at various temperatures.
			(a), (b) XAS and XMCD spectra at the Co $L_{2,3}$ edge. 
			(c), (d) XAS and XMCD spectra at the V $L_{2,3}$ edge.
		}\label{fig:allspc}
	\end{figure}
	Figure \ref{fig:allspc} shows XAS and XMCD spectra of the \sn\ single crystal under a magnetic field of 0.5 T along the [001] and [111] directions at various temperatures. 
	Here, the XMCD spectra have been divided by $\cos \theta_H$ for comparison because the XMCD intensity is proportional to the projection of the magnetization onto the X-ray incident direction. 
	The spectral line shapes of the XAS and XMCD do not show any notable dependence on the magnetic field direction and temperature except for the intensities of XMCD. 
	As shown in Figs. \ref{fig:allspc} (b) and (d), the anisotropy of the XMCD intensity decreases with increasing temperature. This behavior is consistent with bulk magnetization measurements \cite{Koborinai2016,Shimono2016}.

	In order to clarify the electronic state of the excess Co ion and its role in the orbital-glass state,  we have analyzed the obtained Co XAS and XMCD spectra using CI cluster-model calculation \cite{Tanaka1994}.
	Since the $\text{Co}:\text{V}$ ratio of this sample was estimated to be $1.21:1.79$, some of the V$^{3+}$ ions at the octahedral site should be replaced by the excess Co$^{3+}$ ions. 
	In order to take the excess Co$^{3+}$ ions into account, we calculated the spectra of Co both at the tetrahedral site (major component) and the octahedral site (minor component) and the weighted sum of the calculated spectra for the two sites were used for fitting. 
	The ratio of the Co$^{2+}$ ions at the tetrahedral site and the Co$^{3+}$ ions at the octahedral site were fixed at $\text{Co}^{2+}$ $(T_d):\text{Co}^{3+} (O_h) = 1:0.21$. 
	As for the tetrahedral site, the crystal-field splitting 10$Dq$ and the $d$-$d$ Coulomb interaction energy $U_{dd}$ were assumed to be $-$0.5 eV and 5.0 eV, respectively. 
	The charge-transfer energy $\it{\Delta}$ and the Slater-Koster parameter ($pd\sigma$) were treated as adjustable parameters. 
	As for the octahedral site, 10$Dq$ was set to 1.1 eV and 0.7 eV in order to simulate the Co$^{3+}$ low-spin (LS) [$(t_{2g})^{6} (e_g)^0$] and high-spin (HS) [$(t_{2g})^4 (e_g)^2$] states, respectively. 
	Other parameters were adopted from the previous report which reproduced the spectra of Co$^{3+}$ in $O_h$ site of EuCoO$_3$ \cite{Hu2004}.
	\begin{figure}
		\includegraphics[clip, width=8.6cm]{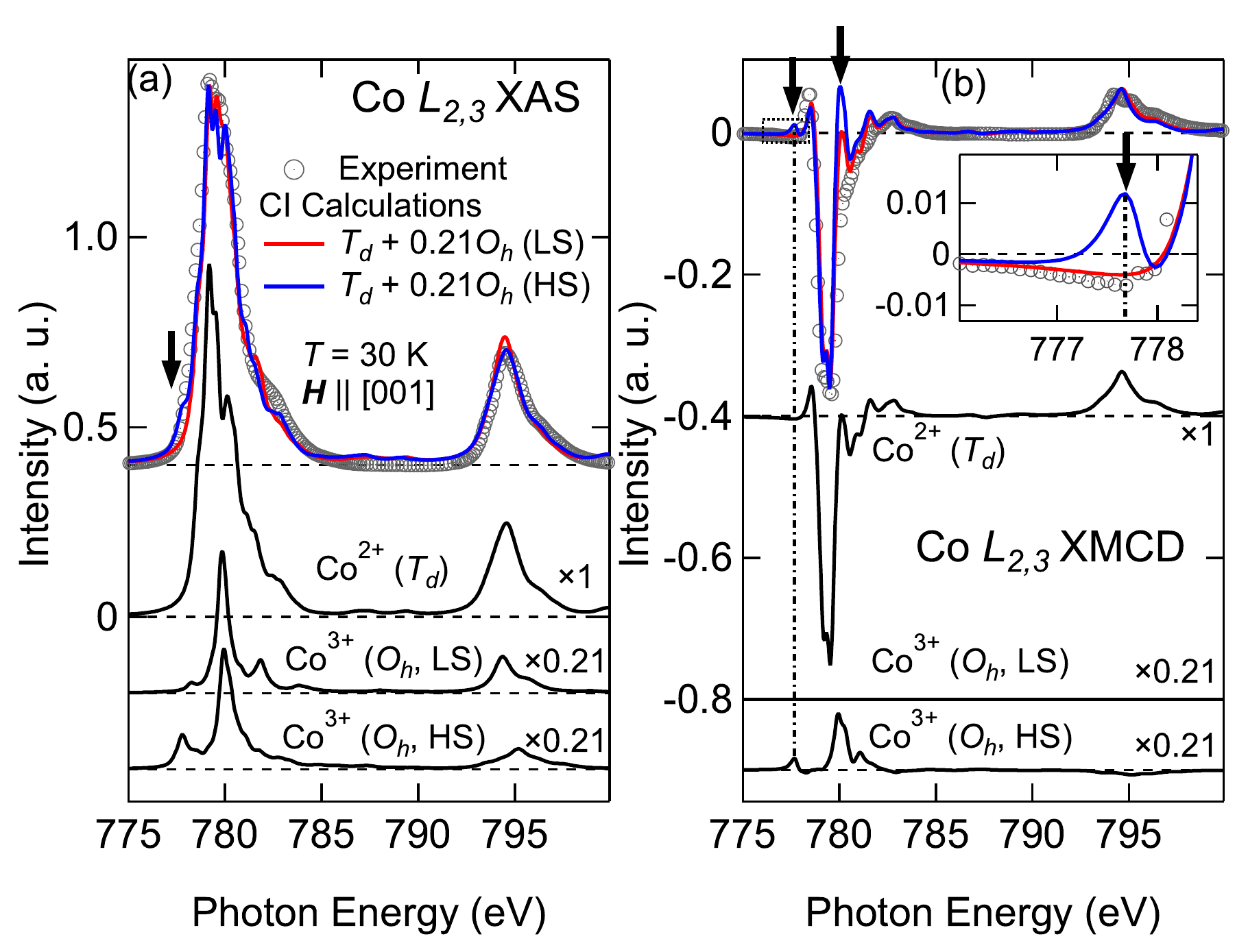}
		\caption{
			Comparison between calculated (solid curves) and experimental (open circles) XAS (a) and XMCD spectra (b) at the Co $L_{2,3}$ edge of \sn. 
			The excess Co ions are assumed to be in the HS (blue curves) and LS (red curves) states. 
			The inset of panel (b) shows a magnified view of the spectra around 776 eV. 
			Discrepancies between the HS calculation and the experiment are marked by black arrows.
		}\label{fig:Co_calc}
	\end{figure}

	Figure \ref{fig:Co_calc} shows comparison between the experimental and calculated Co $L_{2,3}$ XAS and XMCD spectra of \sn. 
	Two calculations were carried out by assuming two different scenarios: One is that the excess Co$^{3+}$ ions take the LS state (red curves), and the other is that they take the HS state (blue curves). 
	The  adjusted parameters for these calculations are listed in Table \ref{table:para}. 
	(The parameters used for the calculation of Co$^{3+}$ ions at the octahedral site are described in the previous text.) 
	It is clear that the LS Co$^{3+}$ scenario shows better agreement with experiment, while the HS Co$^{3+}$ scenario shows some structures (pointed by black arrows) which does not exist in the experimental spectra. 
	Thus, the excess Co ions are more likely to take the magnetically and orbitally inactive LS state and, therefore, they can be dealt as a simple random potential similar to the substituted Al atoms of Mn(V$_{1-x}$Al$_x$)O$_4$ \cite{Adachi2005,Omura2012}. 
	Taking into account the long-range orbital ordering in the low-$x$ region ($0 \le x < 0.15$) of Co$_{1+x}$V$_{2-x}$O$_4$ \cite{Ishibashi2017,Shimono2016}, the origin of the orbital-glass state in this material may be the induced random potential due to the excess LS Co$^{3+}$ ions.

	\begin{figure}
		\includegraphics[clip, width=7cm]{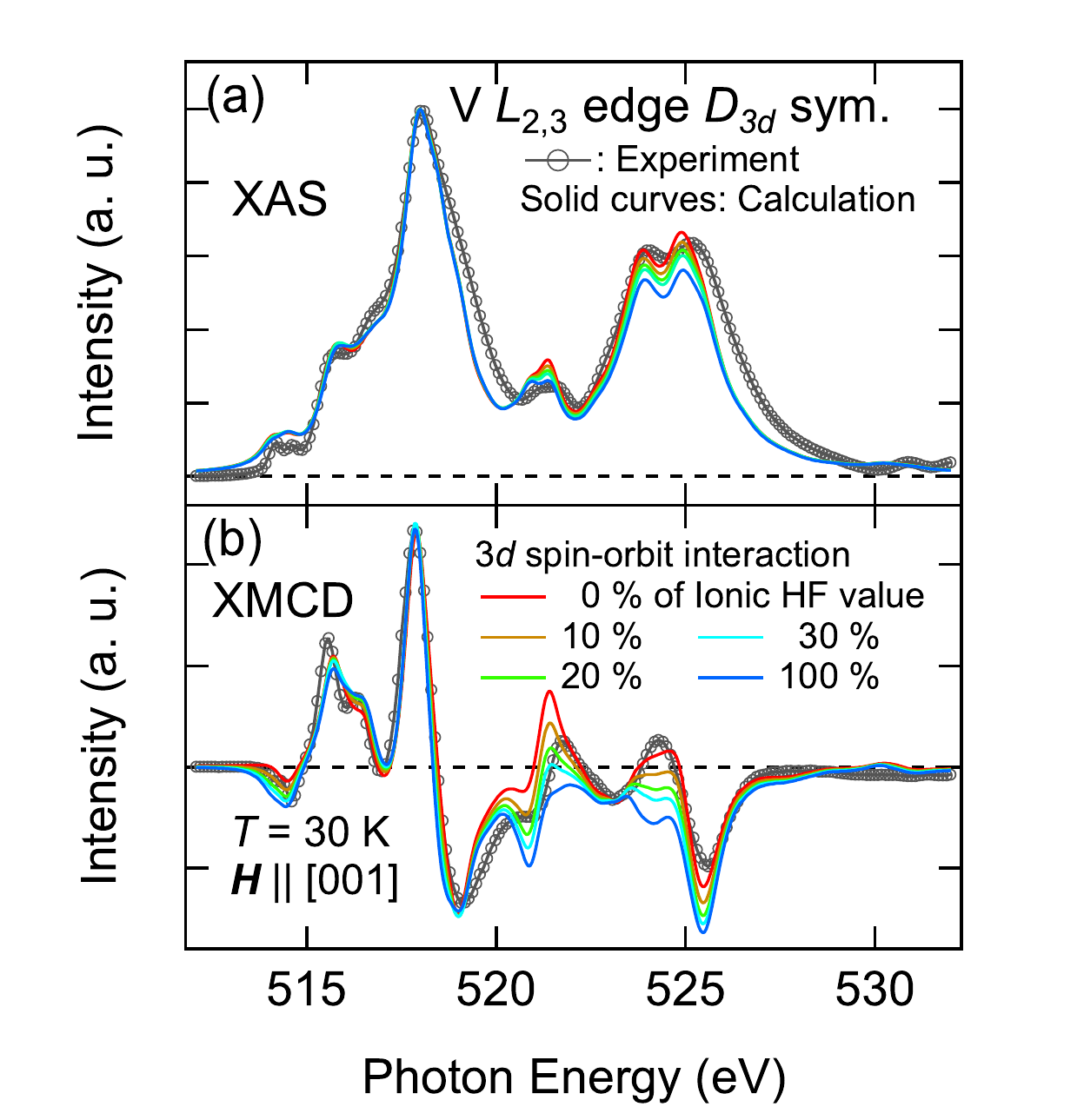}
		\caption{
			Comparison of the calculated and experimental XAS (a) and XMCD (b) spectra at the V $L_{2,3}$ edge of \sn. 
			The trigonal distortion of the $\text{VO}_6$ octahedron is taken into account as the splitting of the $t_{2g}$ level.
			The calculated XAS and XMCD spectra have been normalized to the peak height at 518.0 eV and 517.8 eV, respectively.
		}\label{fig:vsoi}
	\end{figure}

	\begin{table}
		\caption{
			Adjusted parameters in units of eV for the CI cluster-model calculation.
		}\label{table:para}
		\begin{ruledtabular}
		\begin{tabular}{l c c c c}
			Element 	& $\it{\Delta}$ & $pd\sigma$ 	& 10$Dq$ 	& $U_{dd}$	\\ \hline
			Co($T_d$)	& $6.0$			& $1.8$ 		& --- 		& ---		\\ 
			V			& $5.0$			& $2.3$ 		& $0.8$ 	& $5.0$		\\
		\end{tabular}
		\end{ruledtabular}
	\end{table}

	We have also performed CI cluster-model calculations for the V $L_{2,3}$ edge. 
	The parameters used for these calculations are also listed in Table \ref{table:para}.
	Because the important role of strong local trigonal distortion of the VO$_6$ octahedra, which intrinsically exists in the spinel-type structure, has recently been pointed out from first-principles calculations \cite{Sarkar2009a}, we included the trigonal distortion in the cluster-model calculation as a splitting of the $t_{2g}$ level [$\epsilon(e_g') - \epsilon(a_{1g}) = 100 \text{ meV}$].
	As performed in previous reports \cite{Kang2012,Matsuura2015}, we also calculated the spectra for the cases of reduced and full $3d$ spin-orbit interaction (0--30\% and 100\% of the ionic HF values, respectively). 
	Figure \ref{fig:vsoi} shows comparison between the experimental and calculated spectra. 
	The calculation with weak 3\textit{d} spin-orbit interaction reproduced the experiment well even though the trigonal distortion was taken into account.

	\begin{table}
		\caption{
			Spin and orbital magnetic moments in units of $\mu_B$/atom and the total magnetic moments in units of $\mu_B$/f.u. deduced from the XAS and XMCD spectra using the XMCD sum rules and comparison with the calculation. 
			Errors are evaluated from the reproducibility of multiple measurements.
		}\label{table:mom}
		\begin{ruledtabular}
		\begin{tabular}{l c c c}
			Element 	& spin 				& orbital 			& spin + orbital	\\ \hline
			Co($T_d$)	& $2.46 \pm 0.03$ 	& $0.53 \pm 0.03$ 	& $2.99 \pm 0.06$	\\
			V			& $-0.91 \pm 0.05$\footnote{Evaluated by comparison with the calculation (see the main text).} 	& $0.07 \pm 0.07$ 	& $-0.84 \pm 0.12$	\\
			total 		& ---				& ---			 	& $1.48 \pm 0.27$	\\
		\end{tabular}
		\end{ruledtabular}
	\end{table}

	Based on the above results, we have deduced the spin and orbital magnetic moments of Co and V using the XMCD sum rules \cite{Thole1992,Carra1993} and the results of the calculations.
	As for the Co edge, since the excess Co ions take the non-magnetic LS state at the octahedral site, one can adapt the XMCD sum rules by subtracting the contributions of excess Co from the experimental XAS spectrum.
	The relative contributions of the Co$^{2+}$ ions at the tetrahedral site (3 holes in the $3d$ orbitals) and the Co$^{3+}$ ions at the octahedral site (4 holes) have been set as $1:0.28$. 
	Here, the latter contribution is assumed to be $\frac{4}{3}$ times as large as the excess Co concentration estimated from the ICP analysis because the absorption coefficient of an ion is proportional to the number of 3$d$ holes. 
	As for the V edge, since the spin sum rule is not applicable due to the small energy separation betweem the V $2p_{3/2}$ and $2p_{1/2}$ levels \cite{Teramura1996}, the spin magnetic moment was evaluated by comparing the experimental XMCD/XAS intensity ratio with the calculated one.
	The orbital magnetic moment was deduced using the orbital sum rule, where the ionic 3\textit{d}-hele number of 8 (V$^{3+} [3d^2]$) was assumed.
	The deduced spin and orbital magnetic moments are listed in Table \ref{table:mom}. 
	Note that the total magnetic moment is \(M_{Co, Td}+1.79M_{V, Oh}\) because 10.5\% of the octahedral site are occupied by the non-magnetic Co\(^{3+}\) ions.
	We also note that the deduced total magnetic moment $1.48 \pm 0.27 \mu_B$/f.u. is close to the results of bulk SQUID measurements (1.54 $\mu_B$/f.u.). 
	The deduced orbital magnetic moment of V ($0.07 \pm 0.07 \mu_B/$atom) is small ($\ll1\mu_B$).
	Since the ferro-orbital model \cite{Tchernyshyov2004} should have a large orbital magnetic moment, the experimentally observed small one indicates that the ordered orbitals consist of predominantly real-number orbitals such as the antiferro-orbital \cite{Tsunetsugu2003,Motome2004} or orbital-Peierls \cite{Khomskii2005} model.
	
\section{Conclusion}
	The electronic and magnetic properties of \sn, where the V ions are in the orbital-glass state at low temperatures \cite{Koborinai2016,Reig-i-Plessis2016}, have been investigated by XAS and XMCD measurements and subsequent cluster-model analysis. 
	From the line shapes of the XMCD spectra, it was found that the spin magnetic moments of Co are aligned parallel to the applied magnetic field and that those of the V ions are anti-parallel to it. 
	This spin orientation, which is opposite to the expectation from the simple ionic model, is consistent with previous neutron scattering studies. 
	We have performed CI cluster-model calculation for both of the Co and V $L_{2,3}$ edge spectra. 
	As for the Co edge, the calculation assuming the LS state of the excess Co$^{3+}$ ions at the octahedral site well reproduced the experimental spectra.
	This result indicates that the excess Co ions at the octahedral site are in the orbitally and magnetically inactive LS Co$^{3+}$ state, and induce a random potential to the crystal, which is a possible origin of the orbital-glass state in this material.
	The results of the XMCD sum rules are consistent with bulk magnetization measurements.
	The orbital magnetic moment of V was found to be small although finite (0.1 $\mu_B$/atom).
	The results suggest that the ordered orbitals consist of predominantly real-number orbitals.

\section{Acknowledgments}
	Enlightening discussion with Tanusri Saha-Dasgupta and Kartik Samanta is gratefully acknowledged.
	We would like to thank Kenta Amemiya and Masako Sakamaki for valuable technical support at KEK-PF.
	We also thank Masaki Kobayashi for fruitful discussion.
	This work was supported by a Grant-in-Aid for Scientific Research from JSPS (22224005, 15H02109, 15K17696).
	The experiment was done under the approval of the Photon Factory Program Advisory Committee (Proposal No. 2013S2-004, 2016G066, 2016S2-005).
	S.S. acknowledges financial support from Advanced Leading Graduate Course for Photon Science
	(ALPS) and the JSPS Research Fellowship for Young Scientists. 
	Z.C. acknowledges financial support from Materials Education Program for the Futures leaders in Research, Industry and Technology (MERIT). 
	A.F. is an adjunct member of Center for Spintronics Research Network (CSRN), the University of Tokyo, under Spintronics Research Network of Japan (Spin-RNJ).
	Crystal structures were drawn using a \textsc{vesta} software \cite{vesta3}.

\bibliography{newBib.bib}

\end{document}